\documentclass{article}
\usepackage{arxiv}
\usepackage{authblk}
\usepackage[utf8]{inputenc} 
\usepackage[T1]{fontenc}    
\usepackage{url}            
\usepackage{booktabs}       
\usepackage{amsfonts}       
\usepackage{nicefrac}       
\usepackage{microtype}      
\usepackage{lipsum}
\usepackage{amsthm}
\usepackage{amsmath}
\usepackage{bm}
\usepackage{amssymb}
\usepackage{rotating,  graphicx}
\usepackage{cancel}					
\usepackage{changepage}		
\usepackage{pdfpages}		
\usepackage{wrapfig}
\usepackage{tikz}		

    \makeatletter
\def\@fnsymbol#1{\ensuremath{\ifcase#1\or \dagger\or \ddagger\or
   \mathsection\or \mathparagraph\or \|\or **\or \dagger\dagger
   \or \ddagger\ddagger \else\@ctrerr\fi}}

\usepackage{framed}

\usepackage[square,numbers,sort&compress]{natbib}

\usepackage{verbatim} 
\usepackage{siunitx}  
\usepackage{floatrow}
\usepackage{floatrow} 
\usepackage{subfig}
\usepackage[colorlinks=true,linkcolor=blue,citecolor=blue]{hyperref}%
\title{Physical model of end-diastolic and end-systolic pressure-volume relationships of a heart}

\author[1, 2, a]{Yunxiao Zhang}
\author[1, 2, a]{Moritz Kalh\"ofer-K\"ochling}
\author[1, 2, 3, 4]{Eberhard Bodenschatz}
\author[1, 2, *]{Yong Wang}

\affil[1]{\footnotesize Laboratory for Fluid Physics, Pattern Formation and Biocomplexity, Max Planck Institute for Dynamics and Self-Organization, 37077 G\"ottingen, Germany}  
\affil[2]{\footnotesize DZHK (German Center for Cardiovascular Research), Partner Site Göttingen, 37075 G\"ottingen, Germany}
\affil[3]{\footnotesize Institute for Dynamics of Complex Systems, University of G\"ottingen, 37073 G\"ottingen, Germany}
\affil[4]{\footnotesize Laboratory of Atomic and Solid-State Physics and Sibley School of Mechanical and Aerospace Engineering, Cornell University, Ithaca, New York 14853, USA}

\affil[a]{\footnotesize  These authors contributed equally}
\affil[*]{\footnotesize yong.wang@ds.mpg.de}

\begin{document}
\maketitle

\begin{abstract}
Left ventricular (LV) stiffness and contractility, characterized by the end-diastolic and end-systolic pressure-volume relationships (EDPVR \& ESPVR), are two important indicators of the performance of the human heart. Although much research has been conducted on EDPVR and ESPVR, no model with physically interpretable parameters combining both relationships has been presented, thereby impairing the understanding of cardiac physiology and pathology. Here, we present a model that evaluates both EDPVR and ESPVR with physical interpretations of the parameters in a unified framework. Our physics-based model fits the available experimental data and in silico results very well and outperforms existing models. With prescribed parameters, the new model is used to predict the pressure-volume relationships of the left ventricle. Our model provides a deeper understanding of cardiac mechanics and thus will have applications in cardiac research and clinical medicine.
\end{abstract}


\section{Introduction}
A well-functioning heart is critical to the quality of human life \cite{Gilbert2007}. The pump function of the heart can be captured by the pressure-volume (PV) loop, which is a simple and useful framework for analyzing cardiac mechanics from a physical perspective \cite{Witzenburg2017}. Deoxygenated blood is pumped from the right ventricle to the lungs, and in turn, oxygenated blood is pumped from the left ventricle (LV) to the rest of the body. Because the LV is physically subjected to more stress and strain than the right ventricle, the LV is more susceptible to cardiac disease. As a result, cardiac research focuses heavily on the LV. 

\begin{figure}[tb]
\begin{center}
\includegraphics[width=0.6\linewidth]{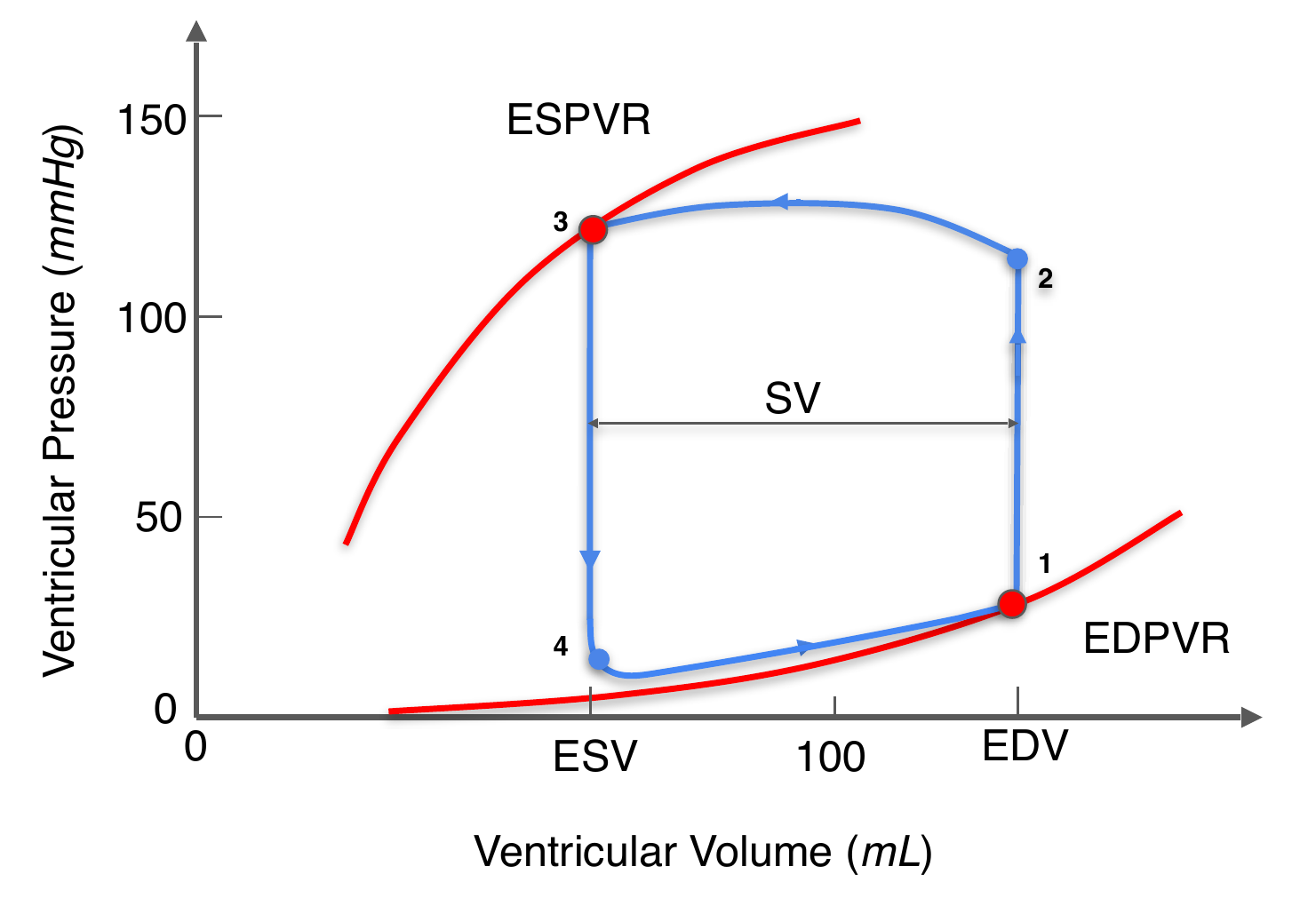}
\caption{\textbf{Illustration of the PV loop, ESPVR, and EDPVR of the LV.} The EDPVR and ESPVR are highlighted in red and the PV loop is in blue. The arrows on the PV loop correspond to the direction of the LV beating cycle. Curve 3-4-1 corresponds to the diastolic phase. At point 1, the ventricular volume reaches its maximum value as the blood fills into the LV, which is called the EDV. Similarly, curve 1-2-3 corresponds to the systolic phase. At point 3, the ventricular volume reaches its minimum value during LV contraction, which is referred to as ESV. The difference between EDV and ESV is the SV, indicating the amount of blood pumped by the LV per cardiac cycle. At different filling pressure and contractility, points 1 and 3 move on a single curve called the EDPVR and ESPVR, respectively. (EDPVR: end-diastolic pressure-volume relationship; ESPVR: end-systolic pressure-volume relationship; EDV: end-diastolic volume; ESV: end-systolic volume; SV: stroke volume.)}
\label{fig:pvloop}
\end{center}
\end{figure}

As shown in Fig. \ref{fig:pvloop}, exemplary for a PV loop, the lower right point (point 1) indicates the end-diastolic (ED) state of the LV.  Varying ED filling pressures yields a change in ED volume. For a heart, these data points fall roughly on a single curve which is termed the end-diastolic pressure-volume relationship (EDPVR) \cite{Sunagawa1983}. The EDPVR is widely used to estimate the mechanical property of myocardium \cite{Brinke2010, Krishnamurthy2013a, Palit2018, Dabiri2018}. The upper left point (point 3) on the PV loop indicates the end-systolic (ES) state of the LV, and the related curve is the end-systolic pressure-volume relationship (ESPVR). The ESPVR and its slope are generally used to describe the contractility of the heart. In this work we present a physics-based model for both the EDPVR and ESPVR.

The EDPVR comprises a number of important markers used by both researchers and clinicians in health assessments. Many studies have shown that the EDPVR has a strong association with heart diseases. Despite its long history, the EDPVR continues to gain increased attention. 
Goto et al. \cite{Goto1985} studied the effects of right ventricular ischemia on LV EDPVR in canine hearts. A leftward and upward shift in the LV EDPVR was observed with no change in LV myocardial performance. 
Brinke et al. \cite{Brinke2010} predicted the LV EDPVR in patients with end-stage heart failure (LV ejection fraction $<40\%$) using single-beat estimation and concluded that such method facilitated less invasive EDPVR estimation.
Schwarzl et al. \cite{Schwarzl2016} showed that, due to LV remodeling, the EDPVR was shifted rightward and leftward in heart failures with reduced ejection fraction and heart failure with preserved ejection fraction, respectively, compared with the reference case. In addition, the risk of heart failure for non-heart failure individuals was found to be associated with the changes in LV capacitance and stiffness, which can be extracted from the EDPVR.
Witzenburg and Holmes \cite{Witzenburg2017} stated that EDPVR contains information not only about the mechanical properties of the myocardium but also about LV geometry. Since cardiac diseases alter the shape or stiffness of the heart and thus the EDPVR, the EDPVR is important and helpful to clinicians.

Despite numerous experimental and clinical studies on the EDPVR, there is relatively little theoretical knowledge, especially on the formulation of the corresponding curves. One commonly used method is to fit the EDPVR to an  exponential form \cite{Artrip2001,Rodriguez2015,Shimizu2018,Burkhoff2005},
\begin{align}
\label{eq:ExponentialModel}
P_{ED} = A(e^{B(V_{ED}-V_0)}-1),
\end{align} 
where $P_{ED}$ is the end-diastolic pressure (EDP); $V_{ED}$ is the end-diastolic volume (EDV);  $A$ and $B$ are fitting parameters; $V_0$ is the reference volume when the ventricular pressure of the LV is zero. The exponential term in Eq. \ref{eq:ExponentialModel} is to reflect the exponential stress-strain relationship of the myocardial mechanical property. Its nonlinearity reflects the fact that diastolic stiffness steadily increases with loading \cite{Brinke2010}. 

Klotz et al. \cite{Klotz2007} suggested that the EDPVR can be non-dimensionalized so that all values for different species, being dog, rat, or human, fall closely on a single curve, called the Klotz curve,
\begin{align}
\label{eq:klotzcurve}
P_{ED} = A_n V_n^{B_n},~\text{with} ~~V_n= \dfrac{V-V_0}{V_{30}-V_{0}},
\end{align}
where $A_n$ and $B_n$ are fitting parameters; $V_n$ is the non-dimensionalized volume; $V_0$ and $V_{30}$ are the ventricular volumes when the ventricular pressures are $\SI{0}{mmHg}$ and $\SI{30}{mmHg}$, respectively. The Klotz curve serves as a reference in some cardiovascular studies. In Nordsletten et al.'s study on human left ventricular diastolic and systolic function \cite{Nordsletten2011}, the Klotz curve served as a reference to validate the numerical data. Hadjicharalambous et al. \cite{Hadjicharalambous2015} took the Klotz curve as a matching target when evaluating the initial parameter set for 3D tagged MRI. Although widely used, the Klotz curve is  an ad-hoc empirical function describing the EDPVR, and does not have physical justification. Furthermore, it shows poor agreement with the experimental data and simulation data at small volumes \cite{Klotz2007, Dabiri2018}.

Besides the exponential model and the Klotz curve, other forms of fitting of the EDPVR can be found in the literature \cite{Burkhoff2005}. These fittings of different orders are more mathematical in nature and do not have sufficient physical implications. Thus, a deeper understanding of the EDPVR and its interaction with myocardial properties and cardiac disease warrants a physical model derived directly from the fundamentals of cardiac mechanics.

Although the EDPVR and ESPVR share common mechanisms, they have mostly been studied separately. Very idealized the ESPVR is assumed to be linear and can be fitted with $P_{ES}=E_{ES}(V_{ES}-V_0)$ \cite{Sagawa1978,Burkhoff2005,Rodriguez2015,Shimizu2018}. Therein $P_{ES}$ and $V_{ES}$ are the pressure and volume at the ES state, respectively; $E_{ES}$ is the slope of the curve, thus the ES elastance. In reality, however, with different contractility, the ESPVR is nonlinear especially over a large volume range \cite{Burkhoff1987,Burkhoff2005,Habigt2021}. Some other fitting functions, such as the bilinear form \cite{Habigt2021} and parabolic form \cite{Burkhoff1987} can also be found in the literature.
Nakano et al. \cite{nakano1990} investigated the nonlinearity of the ESPVR and proposed a contractile index independent of ventricular size. In their work, the LV was mimicked by a thick-walled ellipsoid and the contractile index was used to calculate the wall stress based on the concept of mechanical work. $P_{ES}$ and $V_{ES}$ can then be connected with the relationship between wall stress and thickness. Experiments with 25 healthy dogs showed that the proposed contractile index was independent of ventricular size and geometry.
Sato et al. \cite{Sato1998} measured the ESPVR of rat LV in situ with a catheter and observed contractility dependent nonlinearity in the ESPVR.
Habigt et al. \cite{Habigt2021} focused on the nonlinearity of the ESPVR and investigated the effect of different loading alterations on the shape of the ESPVR in pig hearts. The bilinear behavior of the ESPVR in their experimental data strengthens the argument that the linear model is only a special case of nonlinear ESPVR, which is a strong support for the physics-based ESPVR model with similar nonlinearity that we will present. 
A recent review of invasive analysis for the PV relationships in the LV, including both the EDPVR and ESPVR, can be found in Ref. \cite{Bastos2020}.

Here we present a physics-based model that characterises both the EDPVR and the ESPVR. The model uses parameters derived from the properties of the heart under consideration. The physical properties, such as myocardial stiffness, thickness, and contractility, replace the extensive use of otherwise conjectured fitting parameters found in previous works, as discussed above. Section \ref{SecII} presents the physics-based model.  Section \ref{SecIII} offers a discussion of the model, including its validation and its predictions. Section \ref{SecIV} considers the implications and limitations of the model. Finally, a conclusion is given in Section \ref{SecV}.

\section{The Physics-based Model} \label{SecII}
\subsection{Model Definition and Theory}
\begin{figure}[tb]
\begin{center}
\includegraphics[trim={0 30 0 0}, clip, width=0.75\linewidth]{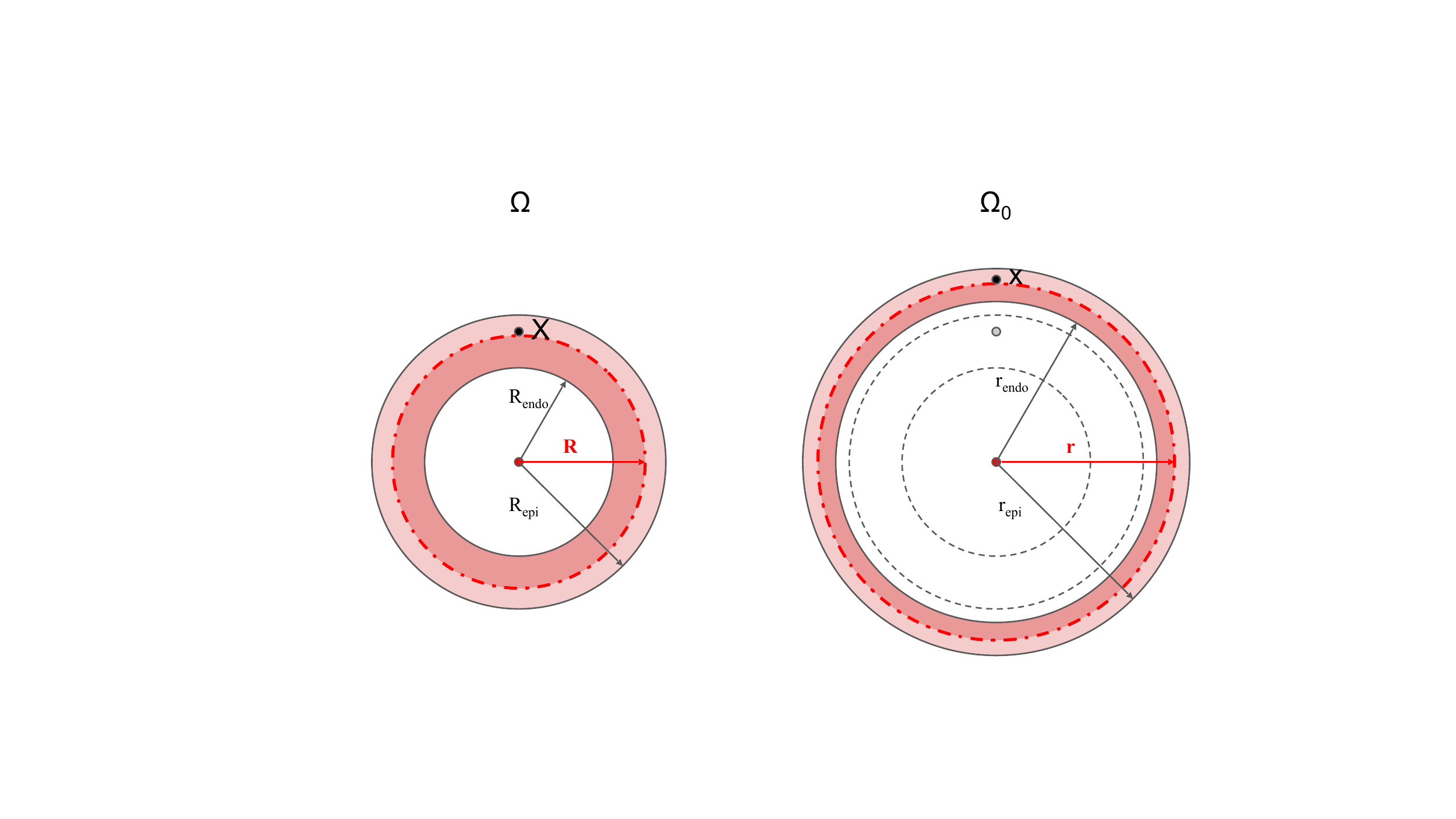}
\caption{\textbf{Schematic diagram of the geometry and its deformation in the physics-based model.} A thick-walled sphere is used to mimic the LV, whose cross section through the sphere center is shown. The reference geometry $\Omega_0$ with inner radius $R_{endo}$ and outer radius $R_{epi}$ is given on the left side. The deformed geometry $\Omega$ with inner radius $r_{endo}$ and outer radius $r_{epi}$ is shown on the right. For any point $\mathbf{X}$ in the reference geometry $\Omega_0$, the corresponding position on the deformed geometry $\Omega$ is $\mathbf{x}$, with radial coordinate $R$ and $r$, respectively. Centrosymmetric deformation is assumed for the model. endo: endocardium; epi: epicardium.}
\label{fig:geometry}
\end{center}
\end{figure}

The schematic of our physics-based model is shown in Fig. \ref{fig:geometry}. Matching the simplicity of the single curve for either EDPVR or ESPVR, the cardiac shape is approximated by a thick-walled sphere. The use of such simplified geometries dates back to the early days of cardiac modeling and can be found still in modern research \cite{Anani2016, MKKThesis}. 
The reference geometry $\Omega_0$ of the LV is shown on the left-hand side, whereas the right-hand side depicts the geometry in a deformed state. The inner and outer radius of the sphere for the reference geometry are $R_{endo}$ and $R_{epi}$, respectively. The wall thickness is thus $R_{epi}-R_{endo}$. While in the deformed geometry, the inner and outer radius become $r_{endo}$ and $r_{epi}$, correspondingly.

A spherical coordinate system is adopted so that its origin is located at the center of the sphere. The three orthogonal basis vectors of coordinate systems are $(\mathbf{e_r, e_{\theta}, e_{\varphi}})$. 
A deformation maps a point $\mathbf{X}$ in the reference geometry $\Omega_0$ to point $\mathbf{x}$ in the deformed geometry $\Omega$. For a given point $\mathbf{X}$ in the reference geometry, the radial coordinate is $R$, while the corresponding radial coordinate of the point $\mathbf{x}$ in the deformed geometry is $r$. The second and third coordinates of the point, $\theta$ and $\varphi$, keep unchanged under deformation due to the assumption of centrosymmetry, which will be clarified later.
The deformation gradient tensor is defined as
\begin{align}
\mathbf{F=\frac{\partial x}{\partial X}}.
\end{align}
Under the spherical coordinate system, it is straightforward to get
\begin{equation}
\mathbf{F} = \text{diag}(\lambda_{\rho},\lambda_{\theta},\lambda_{\varphi}),
\label{eq:f}
\end{equation}
where $\lambda_\rho$ is the radial strain; $\lambda_\theta$ and $\lambda_\varphi$ are two tangential strains. 

We assume that the myocardium is incompressible, resulting in the volumetric strain $J=det(\mathbf{F})=1$, yielding the relation
\begin{equation}
\lambda_{\rho}\lambda_{\theta}\lambda_{\varphi}=1.
\label{eq:j}
\end{equation}
The sphere only has expansion and contraction deformations, which means that the points in the domain only have radial displacement. We have the symmetry constraint
\begin{equation}
\lambda_{\theta}=\lambda_{\varphi}.
\label{eq:sym}
\end{equation}
The strain $\lambda_\theta$ can be calculated by the ratio of the perimeter $l$ of the cross-section through the spherical center (which is represented by the red dotted circle in Fig. \ref{fig:geometry}) on the deformed geometry and $L$ on the reference geometry $\lambda_\theta=l/L$, yielding
\begin{equation}
\label{eq:theta}
\lambda_\theta=\frac{r}{R}.
\end{equation}
Substituting Eqs. \ref{eq:sym} and \ref{eq:theta} into Eq. \ref{eq:j}, we can get the radial strain 
\begin{equation}
\label{eq:lambdaRho}
\lambda_\rho=  \frac{R^2}{r^2}.
\end{equation}

Numerically, the right Cauchy-Green strain tensor $\mathbf{C}$ is a better choice for solving the balance equation than the deformation gradient tensor $\mathbf{F}$, since the former is symmetric and positive definite for all points $\mathbf{X} \in \Omega_0$, which reduces computational costs. Said right Cauchy-Green strain tensor is defined as
\begin{align}
\mathbf{C} = \mathbf{F}^T\mathbf{F}.
\label{eq:c}
\end{align}	
Substituting Eqs. \ref{eq:f} - \ref{eq:sym} into Eq. \ref{eq:c}, the right Cauchy-Green deformation tensor reads
\begin{equation}
\mathbf{C } = \text{diag}\left(\lambda_{\rho}^2,\frac{1}{\lambda_{\rho}},\frac{1}{\lambda_{\rho}}\right).
\label{eq:c2}
\end{equation}
The first invariant of the right Cauchy-Green deformation tensor $I_1$ can be expressed as
\begin{equation}
I_1=\lambda_\rho^2+\frac{2}{\lambda_\rho}.
\end{equation}

Due to the incompressibility of the myocardium, the volume between the inner surface and the red dotted spherical surface (see Fig. \ref{fig:geometry} for reference), stays constant. It follows straightforward that
\begin{align}
\label{eq:v}
R^3-R_\text{endo}^3=r^3-r_\text{endo}^3.
\end{align}
The geometrical parameters are further non-dimensionalized by the inner radius $R_{endo}$ and the ventricular volume $V_0$ of the reference geometry as follows
\begin{align}
\label{eq:non-d}
\hat R =\frac{R}{R_\text{endo}}, \hat{r} =\frac{r}{R_\text{endo}}, \delta =\frac{R-R_\text{endo}}{R_\text{endo}}, 
\Delta=\frac{R_\text{epi}-R_\text{endo}}{R_\text{endo}}, \hat{V}=\frac{V}{V_0},
\end{align}
where $\Delta$ is the non-dimensionalized thickness; $V$ is the ventricular volume at deformed geometry.
With Eqs. \ref{eq:v} and \ref{eq:non-d}, the non-dimensionalized radius can be expressed as
\begin{align}
\label{eq:rnon-dimensionalized}
\hat{r}=\left(\hat{R}^3+\hat{V}-1\right)^{1/3}.
\end{align}

The total elastic energy $W$ stored in the myocardium can be calculated by integrating the energy density function $\Psi$ over the domain $\Omega_0$
\begin{align}
W=\int_{\Omega_0}\Psi \text{d}\Omega.
\label{eq:w}
\end{align}
The sphere experiences an inner pressure $P$, representing the blood pressure inside the LV. While the outer pressure is set to zero, indicating a free boundary condition. Based on classical mechanics, it is known that any mechanical work $W$ performed on the sphere due to a given internal pressure $P$ follows the relation
\begin{align}
P&=\frac{\text{d}W}{\text{d}V}.
\label{eq:p}
\end{align}
Substituting Eqs.\ref{eq:c} - \ref{eq:w} into Eq.\ref{eq:p}, we get an important relationship
\begin{align}
P=-2\int_{0}^{\Delta}\dfrac{\lambda_\rho^2}{\hat{r}}\dfrac{\text{d}\Psi}{\text{d}\lambda_\rho}\text{d}\delta \label{eq:PressureVesselWork}.
\end{align}
Therein, the ventricular pressure is expressed as the integral of the energy density function $\Psi$ over the domain defined by the non-dimensionalized thickness. The EDPVR and ESPVR can be further deducted based on this relationship. Besides such mechanical work approach, another approach based on stress analysis can also be found in Ref. \cite{MKKThesis}.

\subsection{End-diastolic Pressure-volume Relationship}
The myocardium is considered to be a homogeneous, incompressible, anisotropic, and fiber-reinforced soft material that generates active forces. Inspired by experimental data, several constitutive laws (energy density function $\Psi$) for the myocardium have been developed in the last decades, including the orthotropic Holzapfel-Ogden model \cite{Holzapfel2009} and our recently developed squared generalized structure-tensor (SGST) models \cite{Kalhofer-Kochling2020}, in which the fiber dispersion of the myocardium is taken into account. Considering the microstructure of the myocardium, these constitutive laws contain different contributions of isotropic, fibrous, and laminar structures as well as their coupling, and are used for simulations at tissue or organ level \cite{MKKThesis,Chabiniok2016,Baillargeon2014}.

As a first approximation, by neglecting the anisotropy of the cardiac tissue, the isotropic energy function is considered in this work
\begin{align} \label{eq:HOIso} 
\Psi= \dfrac{a}{2b}\left(e^{b(I_1-3)}-1\right), \end{align}
where $a$ and $b$ are mechanical parameters representing the material property, i.e. the stiffness, which can be obtained from experiments. 

\par Incorporating the constitutive law Eq. \ref{eq:HOIso} into Eq. \ref{eq:PressureVesselWork}, one gets the passive contribution of the myocardium on the ventricular pressure, implying the EDPVR
\begin{align}
\boxed{P_{ED}=2a\int_{0}^{\Delta}\dfrac{1-\lambda_\rho^3}{\hat{r}}e^{b(I_1-3)}\text{d}\delta} \label{eq:EDPVR}.
\end{align}
Therein, the input parameters are $a$, $b$ and $\Delta$, while the output is a function indicating the relation between the ventricular volume and the ventricular pressure at the ED state. 
\hspace{1cm}
\subsection{End-systolic Pressure-volume Relationship}
There are two contributions to the stresses and strains in cardiac muscle tissue: the passive and the active components. As for the passive contribution, the tissue generates resistive stress when it is deformed. This tension contributes to the ventricular pressure of the LV. On the other hand, the tissue actively contracts and the active force generated inside the tissue also contributes to the ventricular pressure. These two contributions can be treated as either additive or multiplicative in the constitutive laws. We assume that the two are additive, resulting in the total pressure $P=P_p+P_a$ and the energy function $\Psi = \Psi_p + \Psi_a$. The indices $p$ and $a$ represent the passive and active parts, respectively.

To model the ESPVR, an active contribution, which indicates the active force generated by the myocardium during the ES state, is added onto the passive contribution Eq. \ref{eq:HOIso}. Such an active contribution reads
\begin{equation}
\Psi_a = T_a\left(\frac{\lambda^2}{2}-\lambda_0\lambda\right),
\label{eq:phia1}
\end{equation}
where $T_a$ is the maximum active stress;  $\lambda_0 = l_0 / l_r$ with the sarcomere smallest  length $l_0 = \SI{1.58}{\mu m}$ and rest length $l_r = \SI{1.85}{\mu m}$ \cite{Guccione2001}. The strain of the sarcomere $\lambda$ is defined as
\begin{equation}
\lambda = \sqrt{\mathbf{C:H_a}}.
\label{eq:lambda}
\end{equation}
The active force structure tensor $\mathbf{H_a}$ is defined such that contractile forces act in the tangential plane of the myocardium
\begin{equation}
\mathbf{H_a = I-e_r\otimes e_r},
\label{eq:Ha}
\end{equation}
Substituting Eqs. \ref{eq:lambda} and \ref{eq:Ha} into Eq. \ref{eq:phia1}, the energy function for active force reads
\begin{equation}
\Psi_\text{a} =
T_a\left(\frac{1}{\lambda_{\rho}} - \lambda_0\sqrt{\frac{2}{\lambda_{\rho}}}\right).
\label{eq:phia}
\end{equation}
Incorporating Eq. \ref{eq:phia} into Eq. \ref{eq:PressureVesselWork}, we obtain the active contribution of pressure 
\begin{equation}
P_a = 2T_a\int_0^{\Delta}\frac{1-\lambda_0\sqrt{\frac{\lambda_{\rho}}{2}}}{\hat r}\text{d}\delta .
\label{eq:Pa}
\end{equation}
The pressure at the ES state contains both the passive and active parts. Adding Eq. \ref{eq:Pa} onto Eq. \ref{eq:EDPVR}, we get the ESPVR
\begin{align}
\boxed{P_{ES} =
2\int_{0}^{\Delta}\dfrac{a(1-\lambda_\rho^3)e^{b(I_1-3)}+T_a({1-\lambda_0\sqrt{{\lambda_{\rho}}/{2}}})}{\hat{r}}\text{d}\delta}.
\label{eq:PBESPVR}
\end{align}

\hspace{1cm}

\section{Validation and Discussion} \label{SecIII}
\subsection{Physics-based Model: EDPVR}
\begin{figure}[tb]
\centering
\sidesubfloat[]{\includegraphics[trim={0 0 10 35},clip,width=0.468\textwidth]{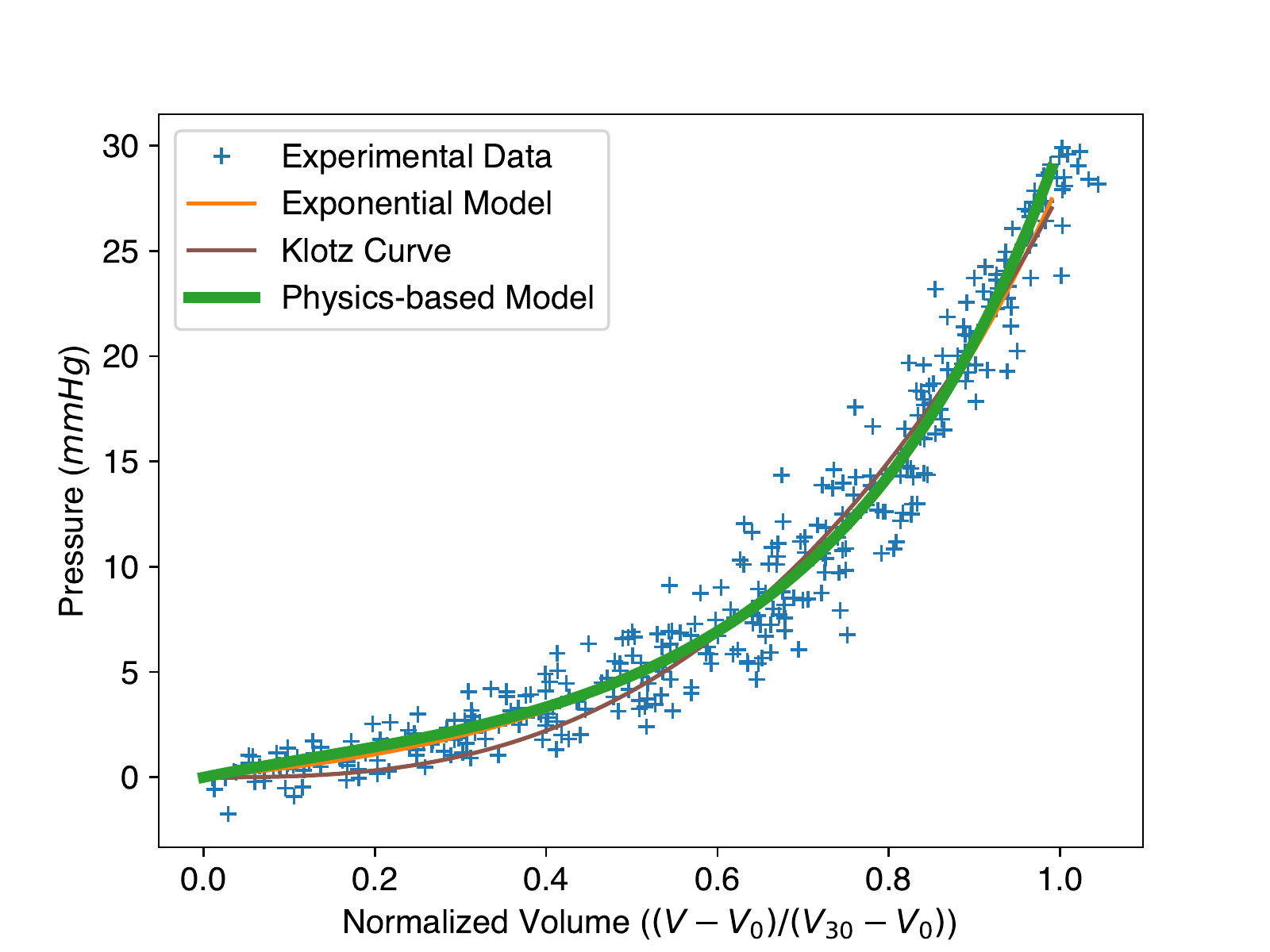}}%
\sidesubfloat[]{\includegraphics[trim={0 0 10 35},clip,width=0.468\textwidth]{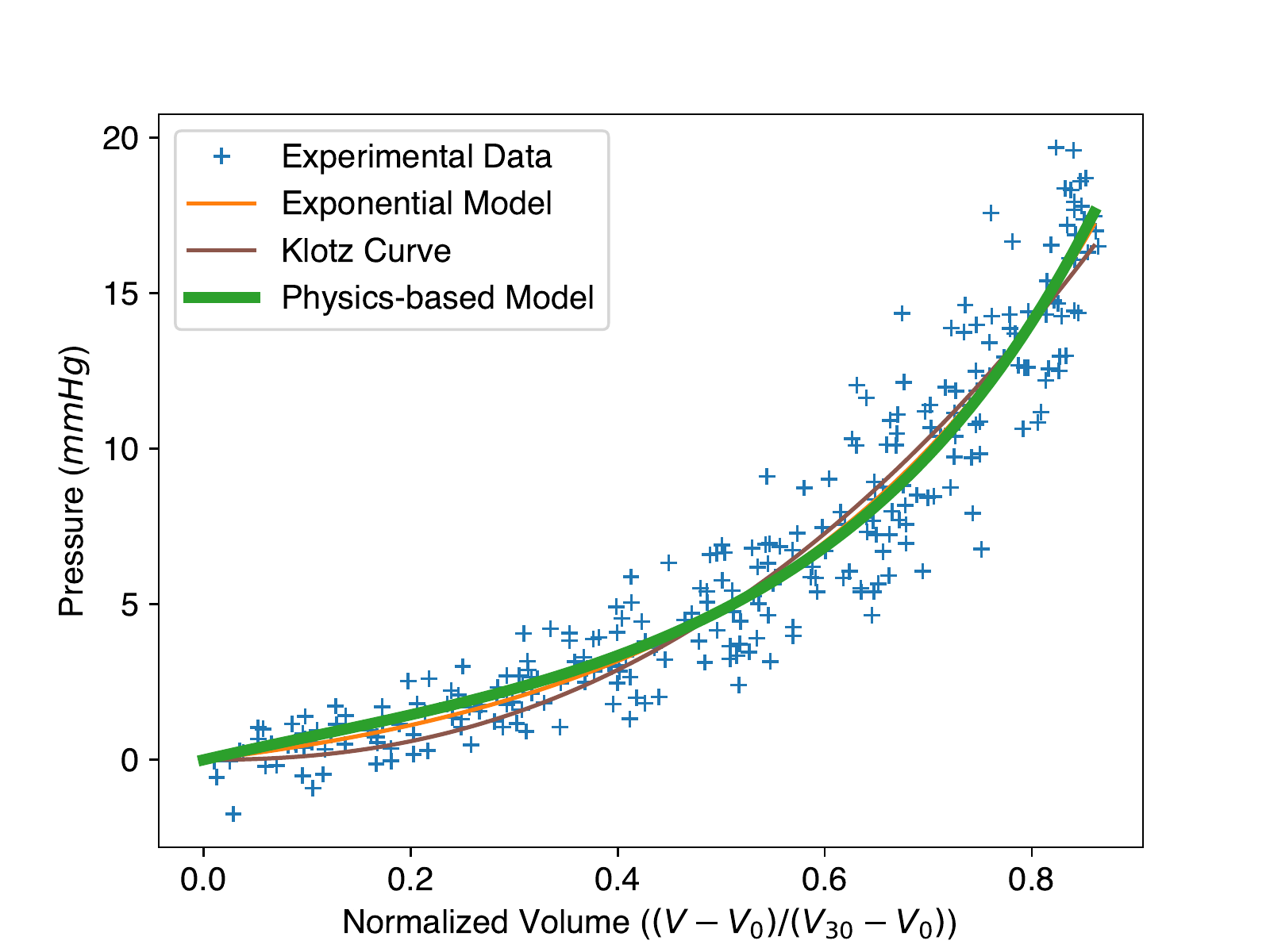}}%
\bigskip

\centering
\sidesubfloat[]{\includegraphics[trim={0 0 10 35},clip,width=0.468\textwidth]{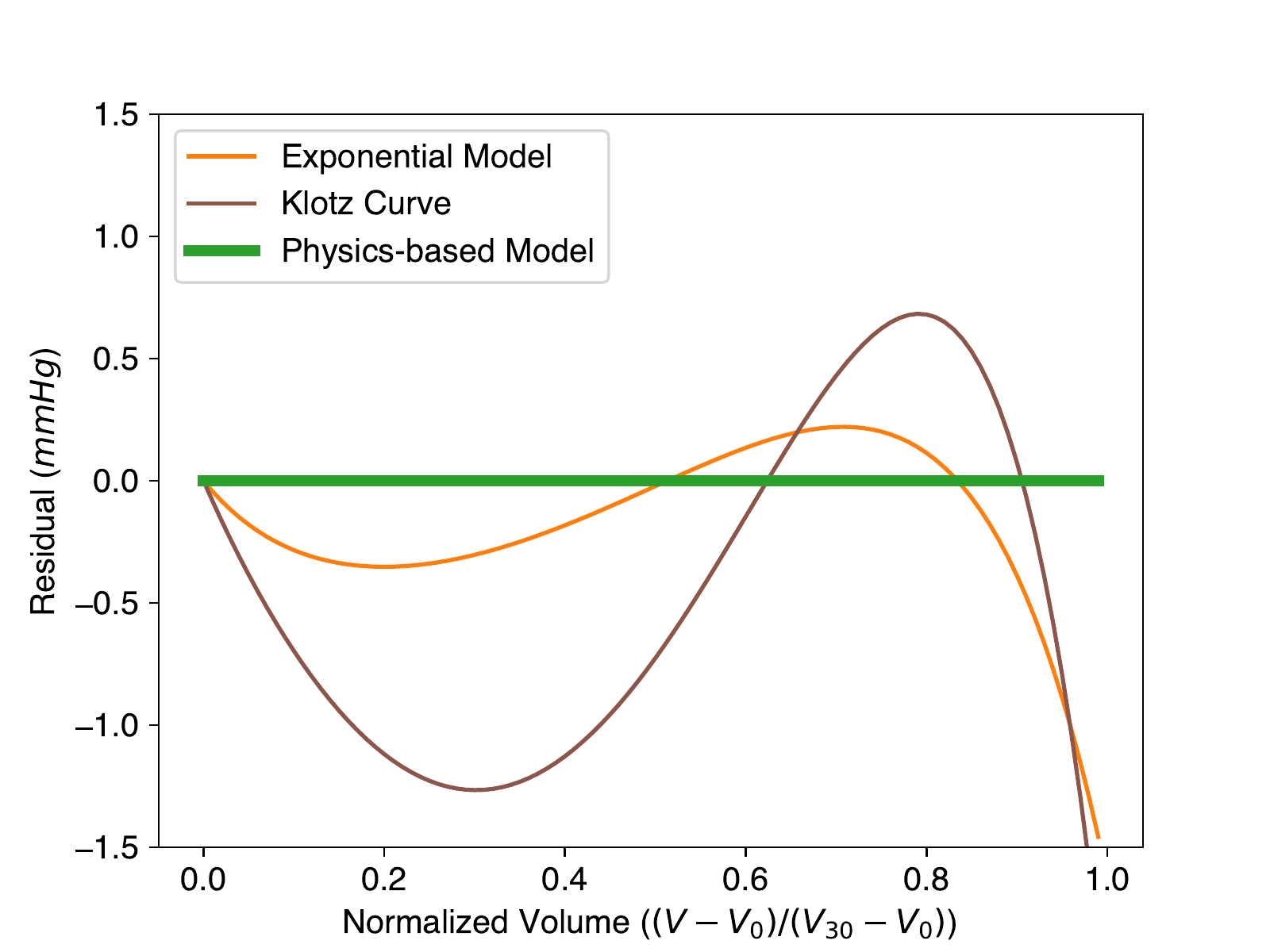}}%
\sidesubfloat[]{\includegraphics[trim={0 0 10 35},clip,width=0.468\textwidth]{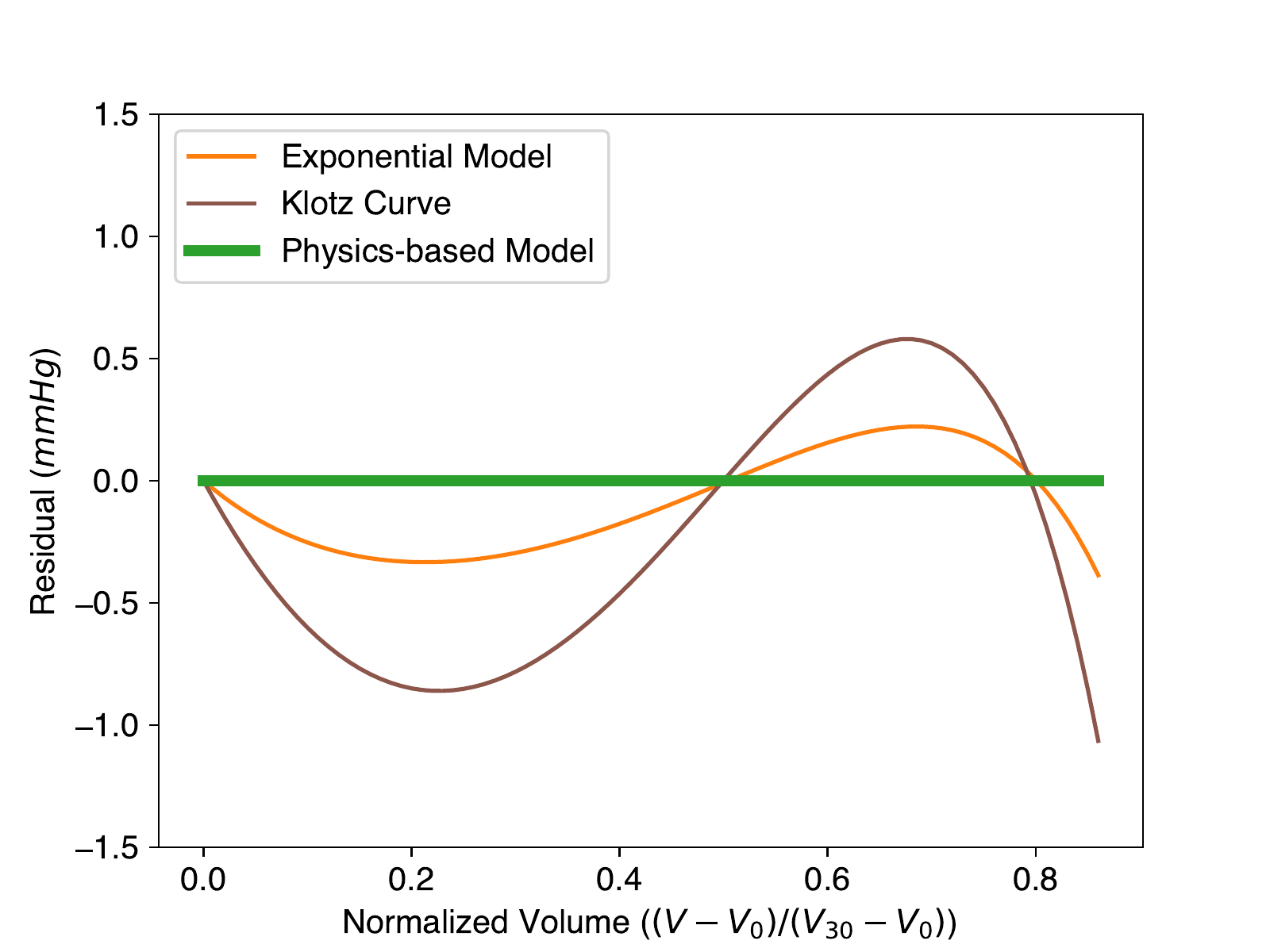}}%

\caption{\textbf{Comparison between the exponential model (Eq. \ref{eq:ExponentialModel}), the Klotz curve (Eq. \ref{eq:klotzcurve}) and the physics-based model (Eq. \ref{eq:EDPVR}) for the EDPVR}. \textbf{(a,c)} The fitted curves to the full experimental dataset and the corresponding residuals. MSEs: exponential model, 2.82; Klotz curve, 3.19; physics-based model, 3.08. \textbf{(b,d)} The fitted curves to the sub-dataset and the corresponding residuals, with pressure no more than \SI{20}{mmHg}. MSEs: exponential model, 2.48; Klotz curve, 2.69; physics-based model, 2.46. Model parameters for subfigures (a,c) and (b,d) are listed in Table \ref{tab:EDPVRFitting}.
\label{fig:double-fitting}} 
\end{figure}

To validate the physics-based EDPVR model, we fit different models to the dataset from Refs. \cite{Klotz2007,Klotz2006} and compare them in Fig. \ref{fig:double-fitting}. The dataset contains ex vivo EDPVR data for 80 human hearts. The fitting was implemented by minimizing the mean squared error (MSE) for the models. The full dataset with pressure up to \SI{30}{mmHg}, and a subset with a physiologically reasonable pressure range (up to \SI{20}{mmHg}), were considered respectively. 

As we can see in Figs. \ref{fig:double-fitting}(a) and (b), the green solid lines representing the physics-based model show a good fit to the experimental data. We compared our model with two other widely used EDPVR models, i.e. the exponential model \cite{Artrip2001,Rodriguez2015,Shimizu2018} and the Klotz curve \cite{Klotz2007}. The Klotz curve (Eq. \ref{eq:klotzcurve}) belongs to the family of polynomial power functions, while the former (Eq. \ref{eq:ExponentialModel}) is classified in the form of exponential functions. 
Our physics-based model entails the combination of an exponential energy function (Eq. \ref{eq:HOIso}) with a volume integral, hence resulting in exponential behaviour. It should be noted that during the curve fitting the original exponential model was adapted to the same non-dimensionalized form as the Klotz curve. This was done by replacing the term $V-V_0$ with $V_n$, which yields $P_{ED} = A(e^{BV_n}-1)$. 

The optimized parameters for the three models are given in Table \ref{tab:EDPVRFitting}. Contrary to previous models, the parameters in our model have a  physical meaning. For example, $a$ and $b$ together reflect the stiffness of the material. The fitted values of $a$ and $b$ of our model are close to most values from experiments and simulations in the literature \cite{Nikou2016, Baillargeon2014, Ziervogel2008}, although parameter estimates themselves often vary considerably across different datasets and experimental protocols. These two parameters are exactly the same as those in the isotropic constitutive law (Eq. \ref{eq:HOIso}). $\Delta$ is the non-dimensionalized thickness of the LV wall, which is an important measure of the ventricular geometry. 

\begin{table*}[b!]
\caption{The least-square fits for the three EDPVR models with respect to the datasets presented in Fig. \ref{fig:double-fitting}. MSE: Mean Squared Error.}
\begin{center}
\begin{tabular}{l|ccc|ccc|cccc}
\hline\hline
&\multicolumn{3}{c|}{Exponential model, Eq. \ref{eq:ExponentialModel}} 
&\multicolumn{3}{c|}{Klotz curve, Eq. \ref{eq:klotzcurve}} 
&\multicolumn{4}{c}{Physics-based model, Eq. \ref{eq:EDPVR}} 
\\
&$A$\,(kPa)	   &$B$	        &MSE 
&$A_n$\,(kPa)	   &$B_n$	        &MSE 
& $\Delta$     & $a$\,(kPa)	 &$b$			&MSE 
\\
\hline
Full dataset &0.16 &3.18 &2.82   &3.70 &2.76 &3.19   &0.27 &1.15 &3.82 &3.08 \\
Sub dataset  &0.18 &3.07 &2.48   &3.10 &2.27 &2.69   &0.27 &2.10 &8.71 &2.46  \\ 
\hline\hline
\end{tabular} 
\label{tab:EDPVRFitting}
\end{center}
\end{table*}

The resulting curves of the three models in the full dataset are shown in Fig. \ref{fig:double-fitting}(a). The MSEs of the original exponential model, the Klotz curve, and the physics-based model are 2.82, 3.19, and 3.08, respectively. Since our physics-based model uses an exponential constitutive law, it is fundamentally similar to the original exponential model. This is why the curves of the physics-based model and the original exponential model are almost identical, and both perform better than the Klotz curve. Fig. \ref{fig:double-fitting}(c) presents the residuals of the exponential model and the Klotz curve, using the physics-based model as a reference. It can be seen that the exponential model is much closer to the physics-based model, especially in the region with small volumes. It is also worth mentioning that the Klotz curve is not consistent with its definition at the maximum volume or pressure. When the non-dimensionalized volume $V_n$ is equal to 1, the resulting pressure in the Klotz curve is the same as the value of $A_n$, which, by design, may be different from the expected \SI{30}{mmHg}.

The physics-based model shows its strong utility  for small volumes. 
The non-dimensionalized volume of this region ranges from 0 to 0.8, corresponding to the pressure 0-\SI{20}{mmHg}, which covers the EDP of the human heart. To better evaluate these three models within this physiologically reasonable range, we generated a sub-dataset with pressures no more than \SI{20}{mmHg}. The results are shown in Figs. \ref{fig:double-fitting}(b) and (d). Here, the physics-based model shows the best fit with a MSE of 2.46. The MSE for the exponential model is 2.48. The Klotz curve has the worst fit with a MSE of 2.69, indicating its weakness at small ventricular volumes.

The Klotz curve is often used as a reference when estimating material parameters of myocardium, like stiffness, in numerical simulations \cite{Krishnamurthy2013,Sack2018,Hadjicharalambous2017}. These simulations mostly use exponential constitutive laws to describe the mechanical properties of the myocardium. Based on the above-mentioned comparison, the new physics-based model shows the capacity to replace the Klotz curve in similar simulations in the future.

\begin{figure}
\centering
\sidesubfloat[]{\includegraphics[trim={0 0 10 35},clip,width=0.468\textwidth]{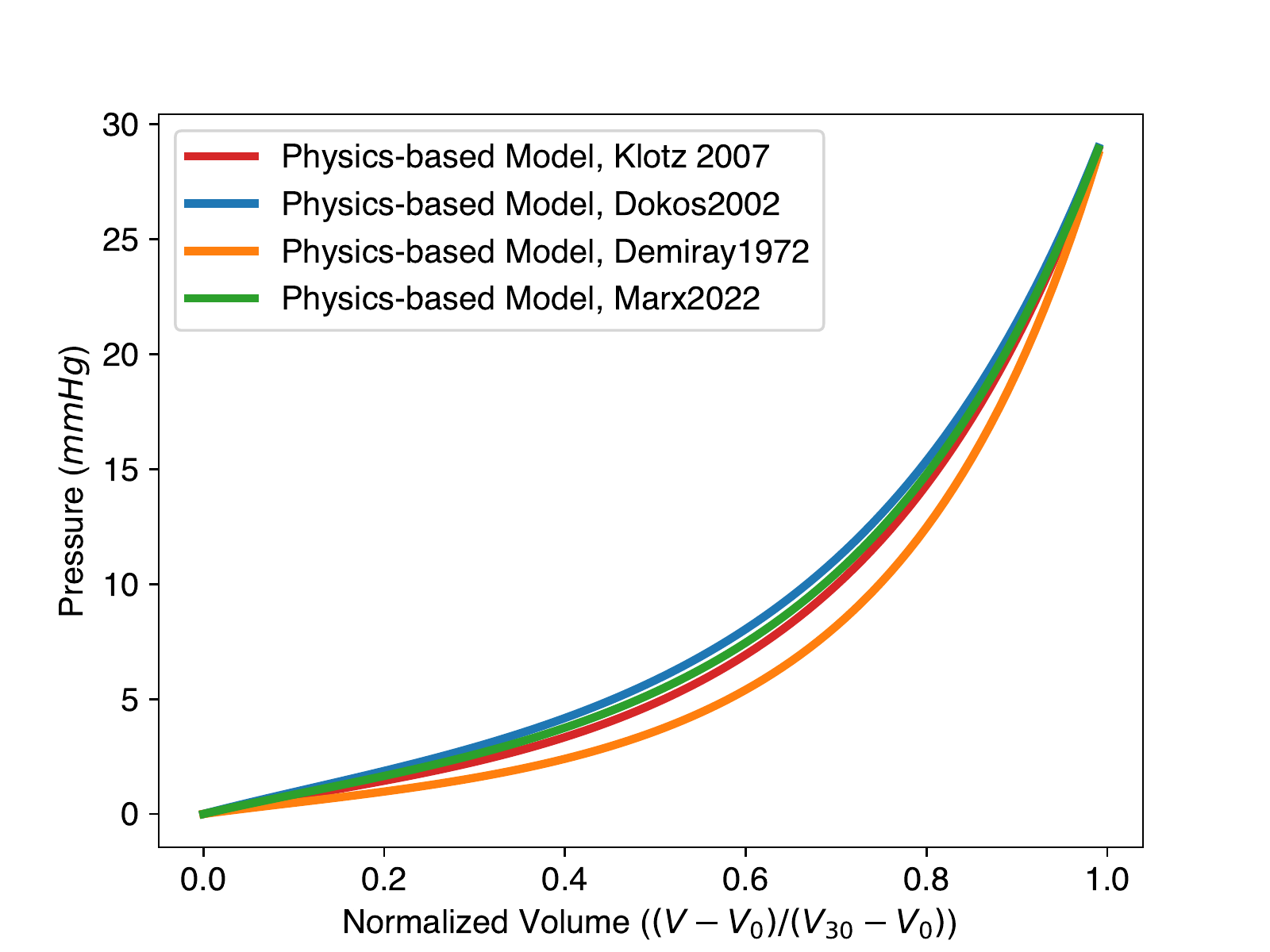}}%
\sidesubfloat[]{\includegraphics[trim={0 0 10 35},clip,width=0.468\textwidth]{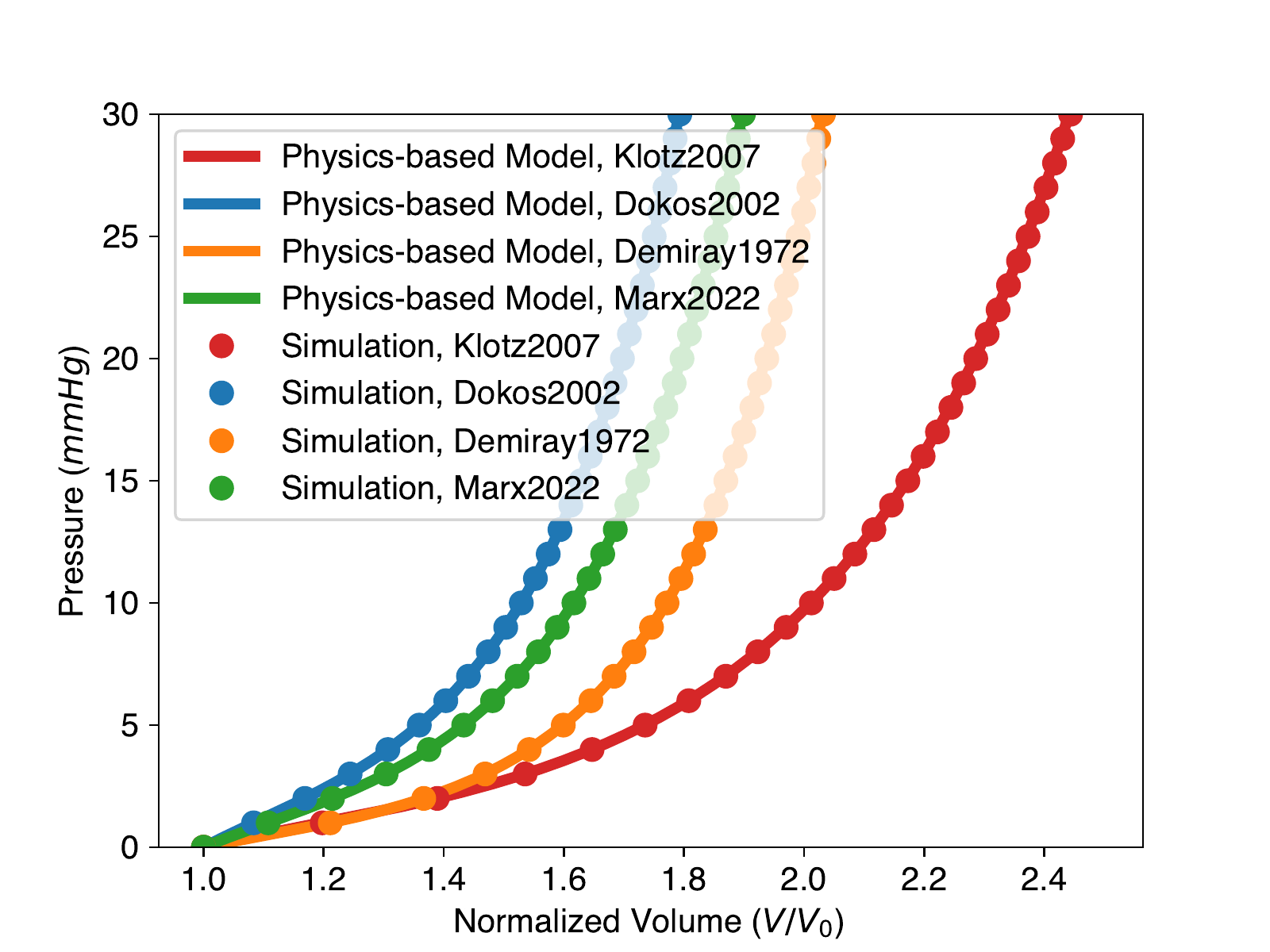}}%
\bigskip

\caption{
\textbf{The physics-based model predicts the effect of varying mechanical properties on the EDPVR with in silico validation.}
\textbf{(a)} The EDPVR curves predicted by the physics-based model with different parameter sets.
\textbf{(b)} In silico validation of the physics-based model. Simulation results agree with the physics-based model very well for each set of parameters. Parameters used in the physics-based model: Klotz2007: $a = 1.15$ kPa, $b = 3.82$; Dokos2002: $a = 2.52$ kPa, $b = 6.79$; Demiray1972: $a = 1.00$ kPa, $b = 6.5$; Marx2022: $a = 1.98$ kPa, $b = 6.19$. For all curves: $\Delta=0.27$.}
\label{fig:edpvr-changeab}
\end{figure}

\begin{figure}[tb]
\centering
\sidesubfloat[]{\includegraphics[trim={0 0 10 35},clip,width=0.468\textwidth]{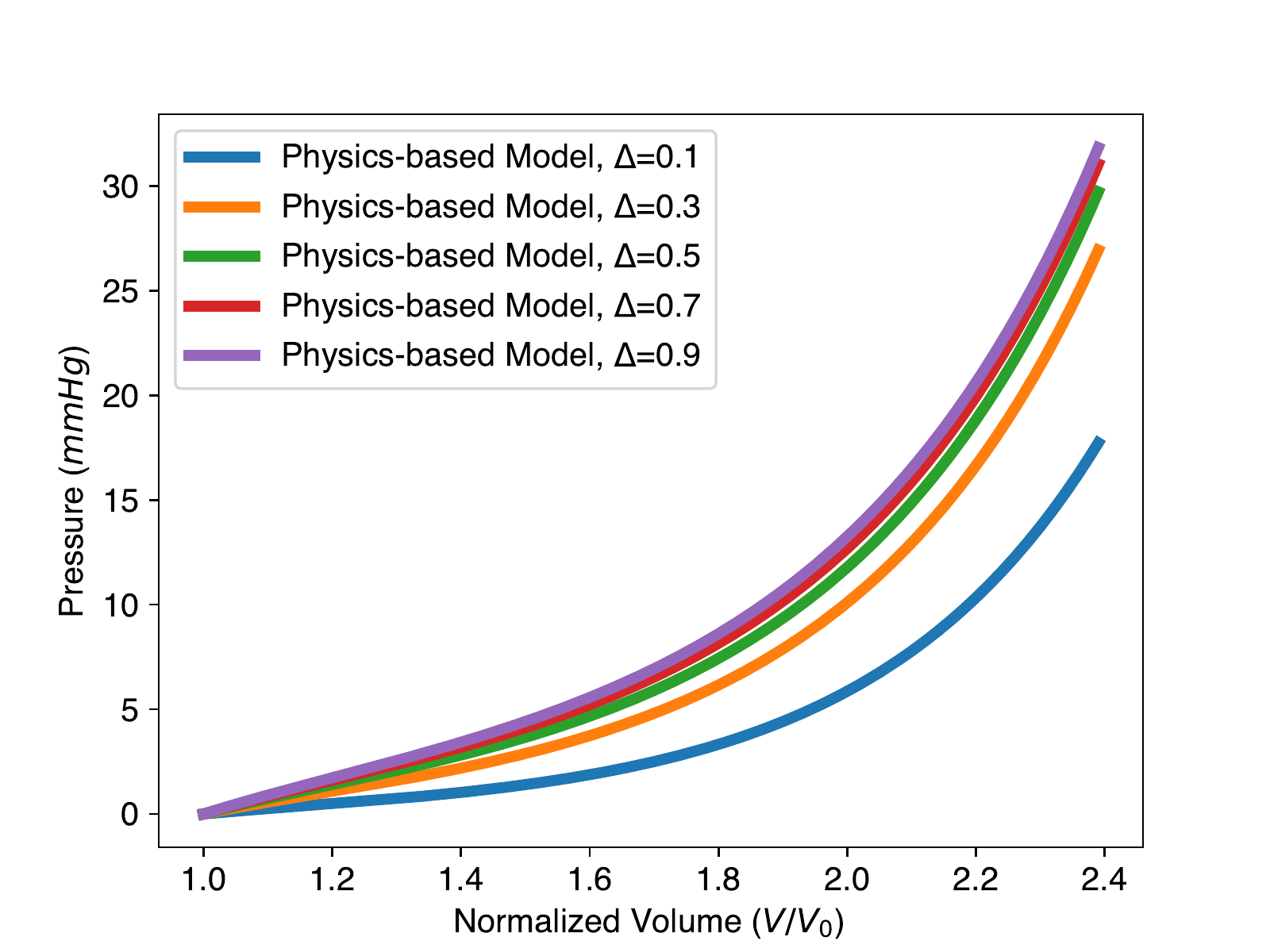}}%
\sidesubfloat[]{\includegraphics[trim={0 0 10 35},clip,width=0.468\textwidth]{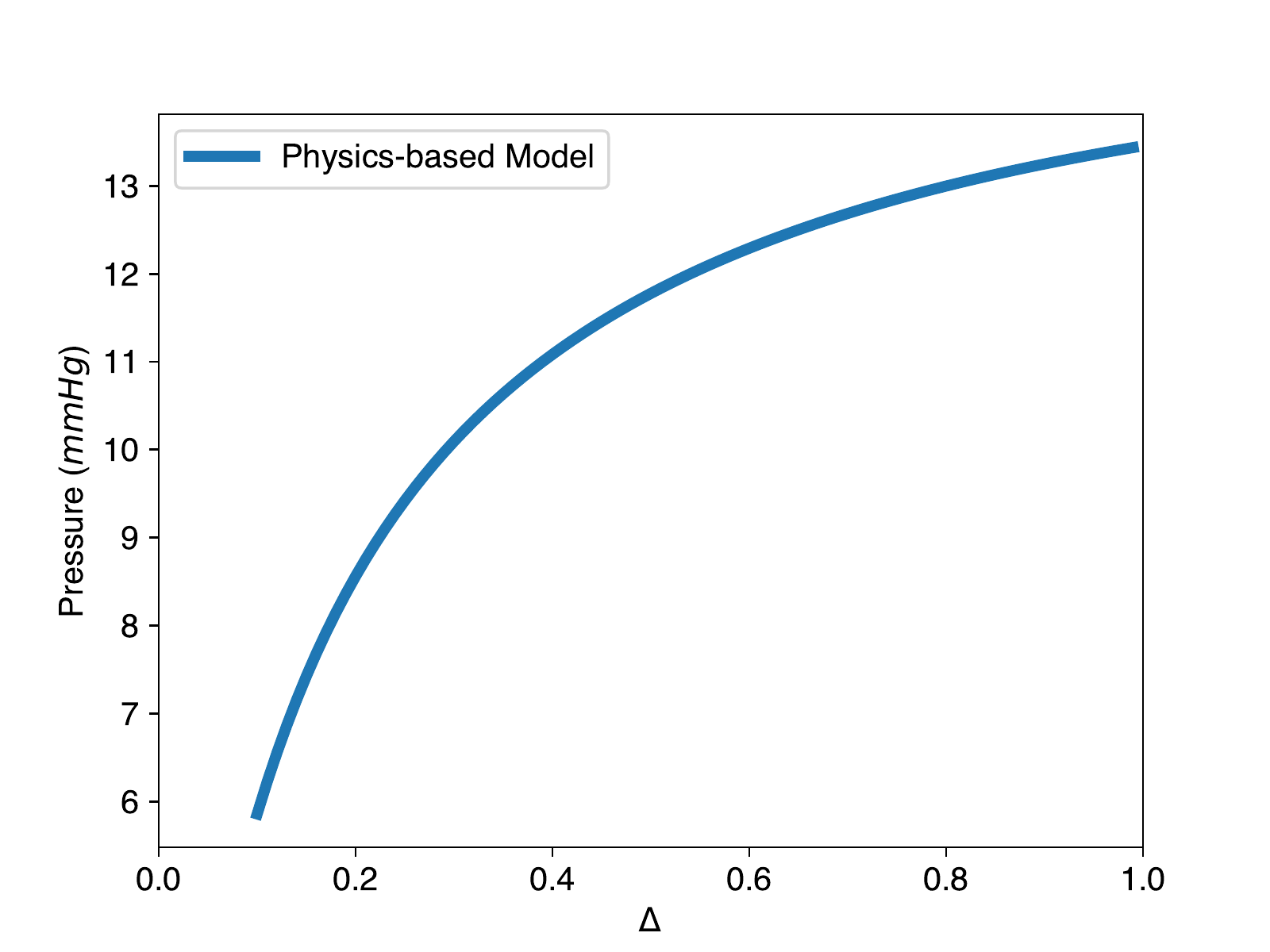}}%
\bigskip

\caption{\textbf{The physics-based model predicts the effect of varying myocardial thickness on the EDPVR.}
\textbf{(a)} The EDPVR curve moves upward with the increase of thickness $\Delta$, while $a$ and $b$ remain constant. 
\textbf{(b)} Nonlinear relationship between the ventricular pressure and the thickness, with constant non-dimensionalized volume 2.0. Parameters used in the physics-based model: $a = 1.15$ kPa and $b = 3.82$.}
\label{fig:edpvr-varying}
\end{figure}

Because of its bottom-up, physical nature, the model can be used to predict the EDPVR of a ventricle with given information, such as mechanical properties and thickness of the myocardium. In Fig. \ref{fig:edpvr-changeab}(a), Dokos2002 \cite{Dokos2002} represents the parameters for pig hearts, Demiray1972 \cite{Demiray1972} and Marx2022 \cite{Marx2022} are for the human hearts. We further performed finite element simulations with the same parameter sets. The curves predicted by the physics-based EDPVR model agree with simulation results excellently, as shown in Fig. \ref{fig:edpvr-changeab}(b). By changing the thickness $\Delta$ in our model, we studied how myocardium thickness affects the EDPVR of the LV, as shown in Fig. \ref{fig:edpvr-varying}(a). An increased myocardium thickness leads to an upward lift of the EDPVR curve. To keep the same volume, higher pressure is needed when increasing the thickness of the LV, as shown in Fig. \ref{fig:edpvr-varying}(b). This reflects the strong utility of a physics-based model over that of simply fit, i.e, the physics-based model is predictive over a wide range of parameters while a fit cannot. 

\subsection{Physics-based Model: ESPVR}
The ESPVR describes the relationship between ventricular pressure and volume at the ES state of the LV as loading conditions change. It is composed of two parts, as shown in Eq. \ref{eq:PBESPVR}. The first part is identical to the EDPVR and the second part comes from the active force generated by the myocardium. The two contributions are shown in Fig. \ref{fig:espvr-contribution}(a). Therein, the solid blue line is the overall ESPVR, while the dashed green one is from the active contraction. A large proportion of the pressure in the LV during the contraction is due to the active force generated by the myocardium. The shape of the ESPVR, especially when the non-dimensionalized ESV is less than 1.0, is mainly determined by the passive resistance due to the deformation of the myocardium. In addition, the ESPVR with small volume is roughly linear, while the overall curve is almost bilinear. This implies that both the linear or bilinear forms of the ESPVR used in the literature have some validity. Please also notice that for the passive part of the ESPVR, the pressure is shown as a negative value. Negative pressure means that the tissue is resisting the deformation caused by its own active contraction. The greater the deformation at ES state, the more negative pressure (resistance stress) is generated, so that the required ventricular pressure is lower. It should be noted that the resistance stress increases in a nonlinear fashion with the decrease of the ESV, due to the nonlinear stress/strain behavior of the tissue. This results in the nonlinear shape of the ESPVR. Our model also shows that the positive slope of the ESPVR is mainly due to the decrease of the resistance stress as ESV increases. 


To validate the proposed ESPVR model, we performed an additional finite element simulation. The geometry of the solid region in the simulation was a sphere with dimensionless inner radius 1.0 and wall thickness 0.27. The inner surface was subjected to a constant pressure mimicking the ventricular pressure from blood, while the outer surface was free. In order to enforce incompressibility of the myocardium, we employed a penalty function $\Psi_v=\kappa(J^2-1)/2-\text{log}(J)$. The bulk modulus $\kappa$ was \SI{1}{GPa}; $J=\text{det(}\mathbf{F})$ was the volumetric strain. The active force generated by the myocardium was determined by the energy function Eq. \ref{eq:phia1}. For the passive response of the myocardium, the energy function was chosen according to Eq. \ref{eq:HOIso}. Parameters used in both the physics-based ESPVR model and numerical simulation are $a = \SI{1.15}{kPa}$, $b = 3.82$, $\lambda_0 = 0.85$, and $T_a = \SI{76.90}{kPa}$. In Fig. \ref{fig:espvr-contribution}(b), the orange dots present simulation results while the blue solid line is from the physics-based ESPVR model. In a large region of the non-dimensionalized volume, the ESPVR predicted by the physics-based model agrees with the simulation results very well.

To further check the validity of our physics-based ESPVR model, we compared it with experimental ESPVR data in Fig. \ref{fig:espvr-exp}. The experimental data was extracted from Ref. \cite{Habigt2021}, by changing afterload pressure. The stress-free volume of the LV is \SI{54}{ml}. The parameters used in our physics-based ESPVR model are: $\Delta = 0.27$, $a=\SI{2.52}{kPa}$, $b=6.79$, $\lambda_0=0.85$, and $T_a= \SI{85}{kPa}$. Our model shows good agreement with the experimental data. 


\begin{figure}[tb]
\centering
\sidesubfloat[]{\includegraphics[trim={0 0 10 35},clip,width=0.468\textwidth]{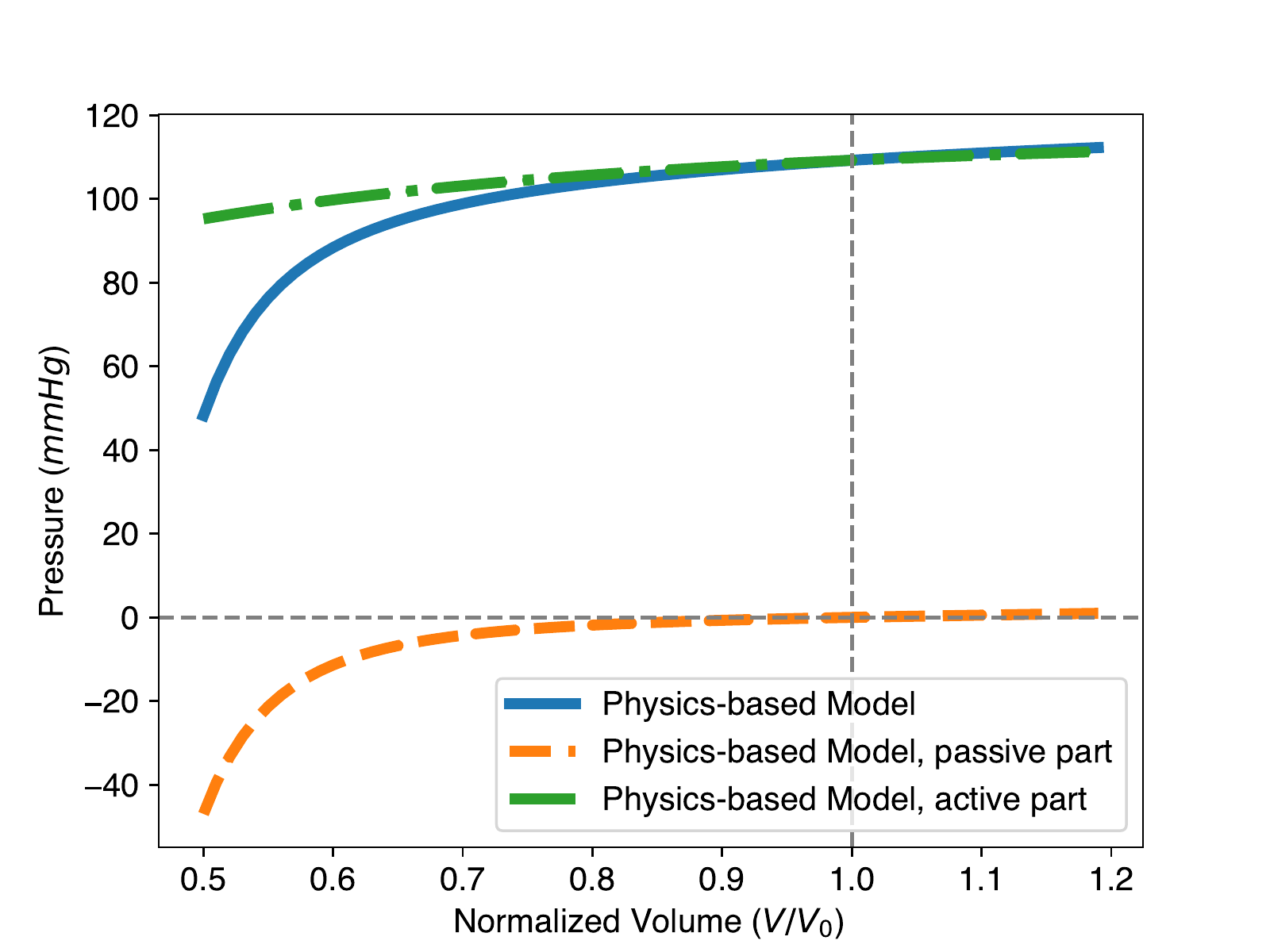}}%
\sidesubfloat[]{\includegraphics[trim={0 0 10 35},clip,width=0.468\textwidth]{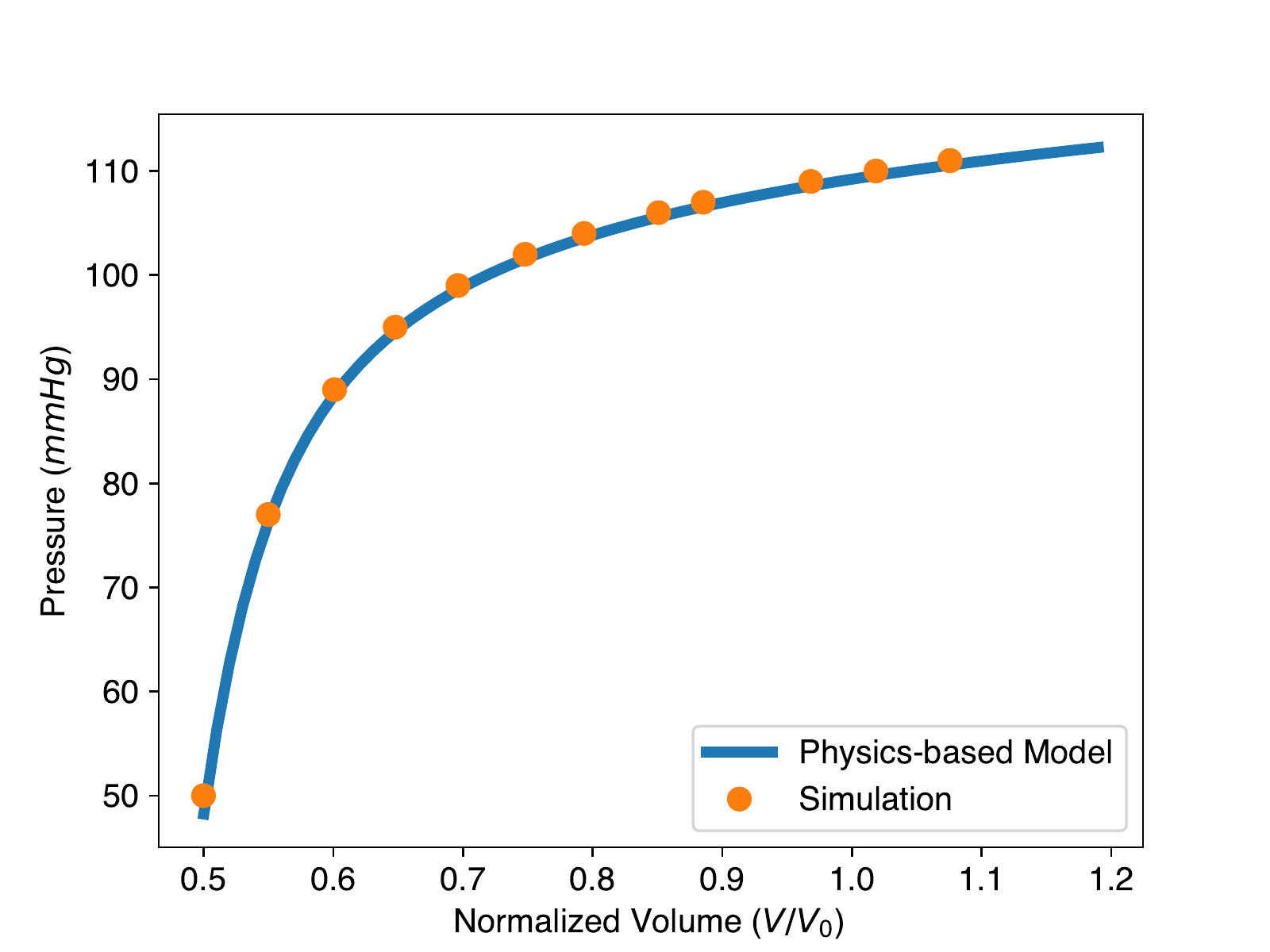}}%
\bigskip

\caption{
\textbf{The intrinsic structure and in silico validation of the physics-based ESPVR model.}
(a) The pressure in the ESPVR has two contributions, one from the active contraction and the other one from the passive resistance of the tissue. The overall level of pressure is mainly determined by the active stress. The shape, especially at the lower ventricular volume region from 0.5 to 0.7, is mostly influenced by passive resistance. (b) The theoretical ESPVR curve fits very well with the simulation results. The parameters for both the theoretical and simulation are $\Delta = 0.27$, $a = \SI{1.15}{kPa}$, $b = 3.82$, $\lambda_0 = 0.85$, and $T_a = \SI{76.9}{kPa}$. }
\label{fig:espvr-contribution}
\end{figure}

\begin{figure}[tb]
\begin{center}
\centering
\includegraphics[width=0.5\textwidth]{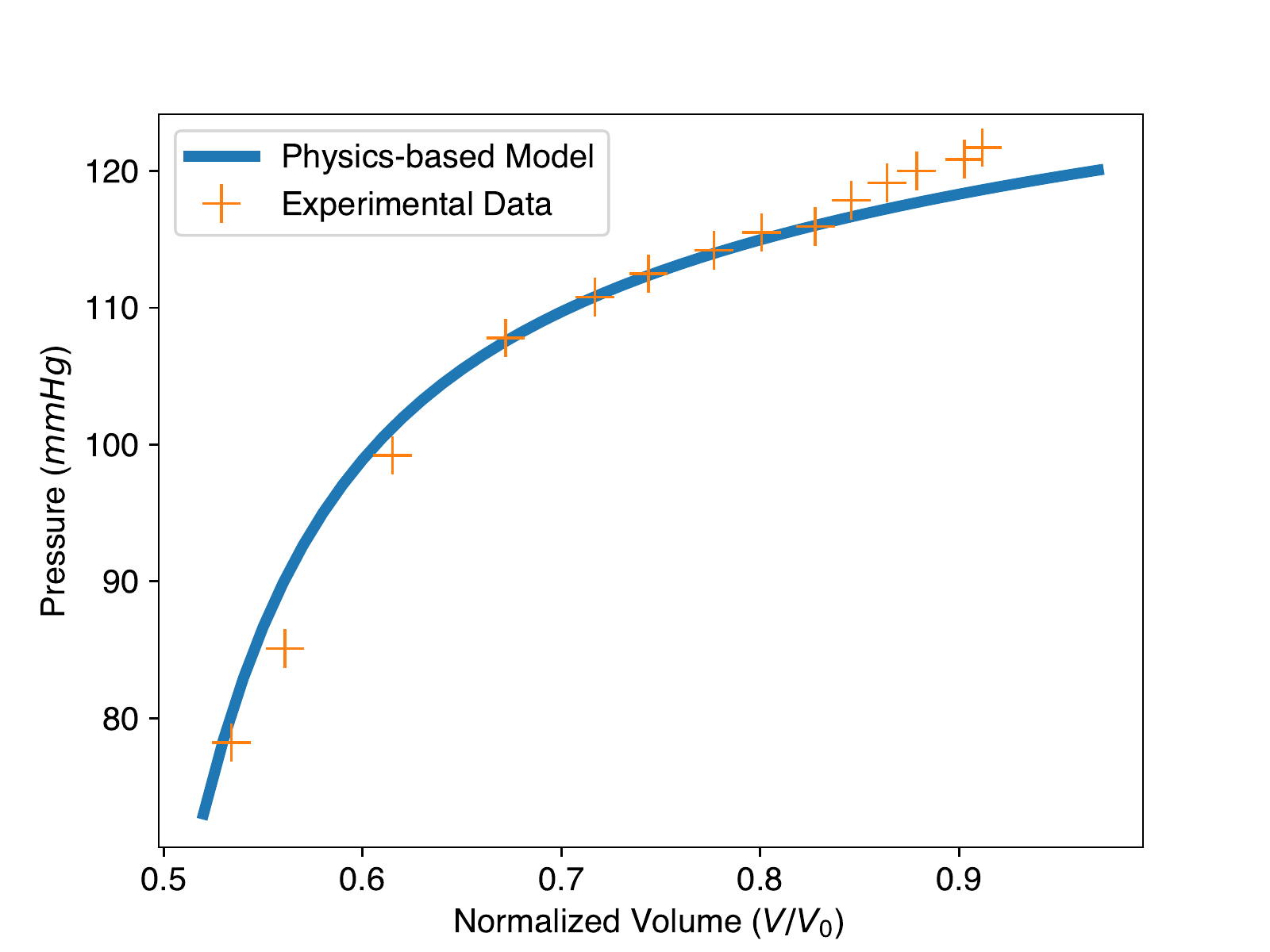}
\caption{\textbf{Comparison of the physics-based model and experimental data for the ESPVR.} The physics-based ESPVR model is shown in blue. The experimental data, shown as orange crosses, is from Ref. \cite{Habigt2021}. The volume of the original data is non-dimensionalized, assuming stress-free volume $V_0=\SI{54}{ml}$.  Parameters used in the physics-based model: $\Delta = 0.27$, $a= \SI{2.52}{kPa}$, $b=6.79$, $\lambda_0=0.85$, and $T_a=\SI{85}{kPa}$. \label{fig:espvr-exp}}
\end{center}
\setcounter{figure}{7}
\end{figure}

\section{Implications and Limitations} \label{SecIV}
Adjusting the parameters of the physics-based model allows the study of left ventricular diseases, such as left ventricular hypertrophy, decreased contractility, and diastolic heart failure. For example, for patients with hypertrophic cardiomyopathy, the heart wall thickens to maintain pump function. This effect is easily visualized with our physics-based model (see Fig. \ref{fig:edpvr-varying}). In addition, the passive parameters $a$ and $b$ can be manipulated to better understand such disease. 

With the physics-based model, one can generate both the EDPVR and ESPVR for the same LV in the same plot, as shown in Fig. \ref{fig:espvr-edpvr}. With given information such as pressure or volume in the ED and ES states, the PV loop will be determined. Therefore, indicators of pump function, such as the stroke volume (SV) and ejection fraction (SV/EDV) can also be calculated. Furthermore, considering new therapies for heart failure, such as engineered muscle tissue \cite{Tiburcy2017,Zimmermann2006}, Fig. \ref{fig:espvr-edpvr} also shows that increasing wall thickness by implanting contractile tissue patch will increase the pump function of the LV. This is because when the wall thickness is increased, the EDPVR hardly changes, while the ESPVR is lifted to the upper left.

\begin{figure}[tb]
\begin{center}
\centering
\includegraphics[width=0.5\textwidth]{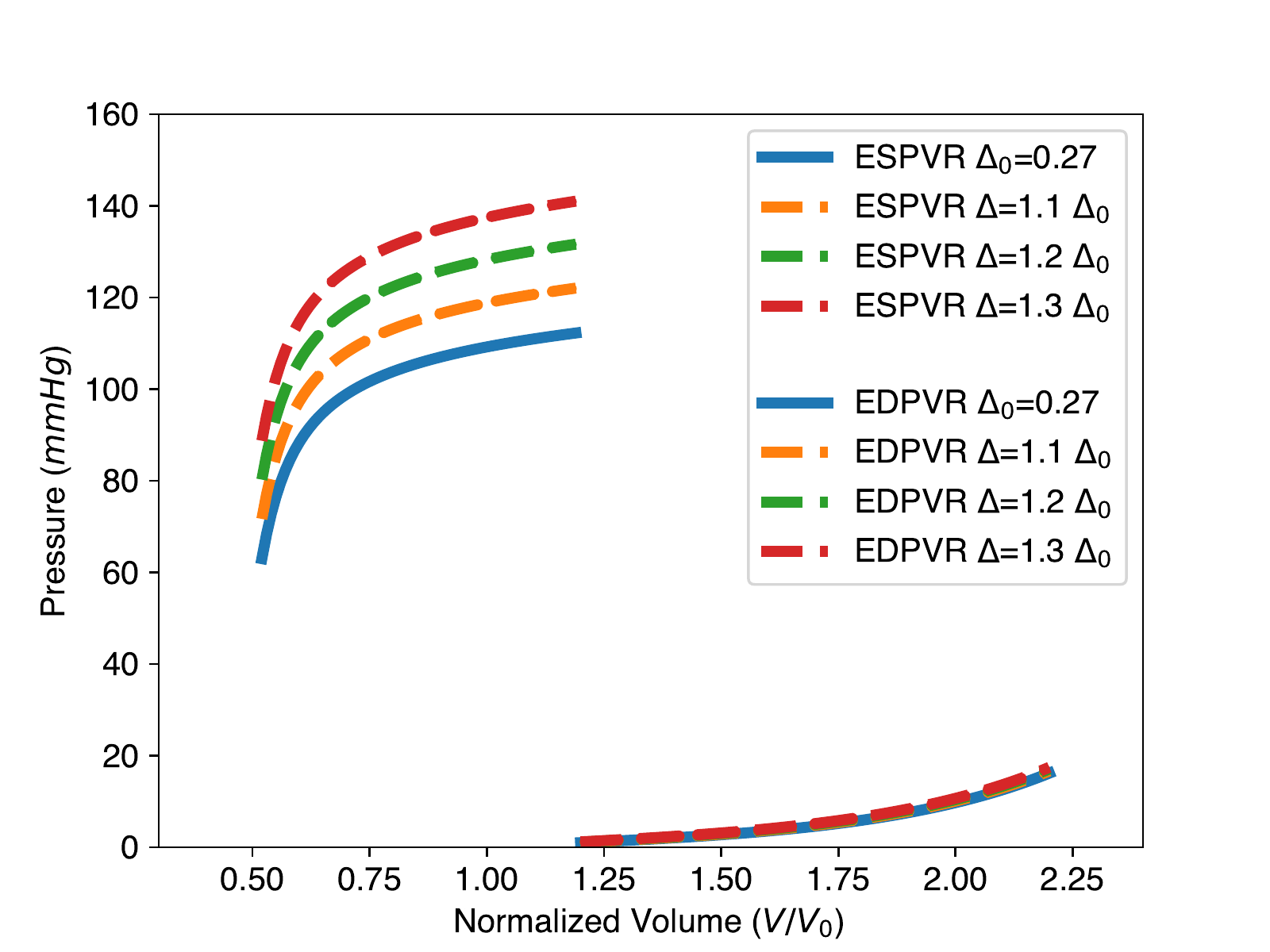}\caption{\textbf{Effect of wall thickness on the EDPVR and ESPVR predicted by the physics-based model.} $\Delta_0=0.27$ is the LV wall thickness of a healthy human heart and is therefore considered here as a reference case. The change in wall thickness has a significant effect on the ESPVR, but a relatively small effect on the EDPVR. Parameters used in the physics-based model: $a= \SI{1.15}{kPa}$, $b=3.82$, $\lambda_0=0.85$, and $T_a=\SI{76.9}{kPa}$. \label{fig:espvr-edpvr}}
\end{center}
\setcounter{figure}{8}
\end{figure}

Our physics-based model provides even more latitude by combining pressure, volume, shape, active force, and mechanical properties of the LV into a unified framework. We notice that 3D printed artificial hearts \cite{Lee2019,Noor2019} or ventricles \cite{Kolawole2021} are attracting increased attention recently, bringing new opportunities for the treatment of heart diseases. The physics-based model proposed in this work can be used to guide 3D printing. For example, with certain mechanical properties and the targeted pump function, our model can predict the required thickness of the heart chamber. For a given ventricular pressure $p$ for which the artificial heart is designed to experience, there is a threshold that the active force $T_a$ must exceed, which can also be obtained from our model.

Furthermore, considering the dynamic cardiac cycle, the ratio of active force $T_a$ to ventricular pressure $p$ gradually increases from diastole to systole. When this ratio exceeds a certain threshold, the LV starts to contract, which means that the volume of the LV is smaller than the stress-free volume $V_0$. This stress-free or pressure-free geometry is widely used in heart simulations. Our model shows that this threshold is only related to the LV thickness and with this value one can obtain the stress-free volume and the associated reference time.

Last but not least, by reducing $b$ to zero, the model reduces to a rubber spherical shell of Neo-Hookean's material. The elastic instabilities ~\cite{Anssari-Benam2022} of spherical inflation can also be reproduced with our model, revealing possible applications of our model beyond the heart.

We are aware that our reductionistic approach cannot fully describe cardiac mechanics. Yet, this simplicity makes it a powerful tool to support comprehensive simulations and diagnostics. The ventricular geometry is not perfectly spherical and, whereas the myocardium is layered and fiber-reinforced soft matter with rotated and dispersed fibers \cite{Sommer2015,NV2017,Khalique2020,Reichardt2020,Kalhofer-Kochling2020}. If all these factors are to be considered, the complexity of the model increases significantly. For such a complex situation, it is recommended to use numerical simulations rather than theoretical models. If the overall pump function and the pressure-volume relationship of the left ventricle are to be considered only, the current simplifications are believed sufficient.

An easy-to-use Python code, for both the physics-based EDPVR and ESPVR, is provided as Supplementary Material.

\section{Conclusions} \label{SecV}
To conclude, we proposed a bottom-up physics-based model incorporating the EDPVR and ESPVR of the LV. The two contributions in this model show the sources of pressure in the end-diastolic and end-systolic states. The model fits existing experimental data well and shows good agreement with simulation results. The model has been shown to be suitable for evaluating LV stiffness and contractility. Conversely, the model can predict the EDPVR and ESPVR of the LV based on the parametric and geometric information of the myocardium. It can also be used to study the relationship between the mechanical properties of the LV and its pump function. The proposed model might provide insight into the study of cardiac mechanisms and be used in clinical medicine.

\section*{Author contributions}
YXZ and MKK contributed equally. MKK proposed the model for the EDPVR. YXZ proposed the model for the ESPVR. YXZ, MKK and YW performed data analysis, and drafted the manuscript. YXZ, MKK, EB and YW reviewed the manuscript.

\section*{Acknowledgements}
This work was supported by the Max Planck Society and the German Center for Cardiovascular Research. We thank Wolfram Zimmermann and Tim Meyer for their continuous support and inspiring scientific discussions.

\section*{Supplementary material}
A code of the physics-based model is provided online.

\bibliographystyle{unsrt} 
\bibliography{references}

\end{document}